\documentclass{aastex}
\usepackage{spr-astr-addons,graphicx}

\RequirePackage{color}

\newcommand{\hab}{HAeBe}
\newcommand{\emaila}{gduchene@berkeley.edu}

\begin{document}

\title{Herbig\,AeBe stars: Multiplicity and consequences}
\slugcomment{Not to appear in Nonlearned J., 45.}
\shorttitle{Herbig\,AeBe stars multiplicity}
\shortauthors{G. Duch\^ene}

\author{G. Duch\^ene\altaffilmark{1}}
\affil{Astronomy Department, University of California, Berkeley, CA 94720-3411 USA}
\email{\emaila}

\altaffiltext{1}{Univ. Grenoble Alpes, IPAG, F-38000 Grenoble, France\\
CNRS, IPAG, F-38000 Grenoble, France}

\begin{abstract}
By virtue of their young age and intermediate mass, Herbig\,AeBe stars represent a cornerstone for our understanding of the mass-dependency of both the stellar and planetary formation processes. In this contribution, I review the current state-of-the-art multiplicity surveys of Herbig\,AeBe stars to assess both the overall frequency of companions and the distribution of key orbital parameters (separation, mass ratio and eccentricity). In a second part, I focus on the interplay between the multiplicity of Herbig\,AeBe stars and the presence and properties of their protoplanetary disks. Overall, it appears that both star and planet formation in the context of intermediate-mass stars proceeds following similar mechanisms as lower-mass stars.
\end{abstract}

\keywords{Binaries: general; stars: early-type; stars: pre-main sequence}

\section{Introduction}

Herbig\,AeBe (\hab) stars are young ($\lesssim 10$\,Myr), intermediate-mass (1.5--8\,$M_\odot$) stars whose defining characteristic is to host circumstellar protoplanetary disks \citep{herbig60, hillenbrand92}. As such, they provide an important perspective on the physics of stellar and planet formation. For one, they are higher mass counterparts to the well-studied Pre-Main Sequence (PMS) T\,Tauri stars (TTS). At the same time, they represent the initial stage of the formation of planetary systems around intermediate-mass stars, which have come in focus in recent years as it has become clear that gas giant planets are even more common around intermediate-mass stars than they are around solar-type stars \citep[e.g.,][]{johnson10}.

It has long been known that stellar multiplicity is an ubiquitous phenomenon that is established during the star formation process itself \citep{mathieu94, duchene13, reipurth14}. Furthermore, there is a strong correlation between the multiplicity frequency and stellar mass on the Main Sequence (MS), so that single stars are a rare occurrence among field intermediate-mass stars \citep{abt83}. The naive expectation is therefore that the multiplicity frequency of \hab\ stars is high. Indeed, the general population of intermediate-mass, mostly diskless, stars in the Sco-Cen OB association has a high multiplicity rate \citep{kouwenhoven07, rizzuto13}. Among \hab\ stars, a high frequency of companions\footnote{Throughout this paper, the frequency of companions refers to the average number of companions per target, defined as $CF=\frac{N_2 + 2 N_3 + \dots}{N_1 + N_2 + N_3 + \dots}$, where $N_1$, $N_2$, $N_3$, ..., represent to the number of single, binary, triple, ... systems, respectively. In the presence of many high-order systems, this quantity can exceed 100\%.} has been found by a number of surveys in the past two decades \citep[e.g.,][]{leinert97, corporon99, baines06}, albeit each within limited detectability ranges and possibly biased samples. Offering an updated view of this topic is one of the main goals of this contribution.

The presence of a close stellar companion can have serious implications on the formation of planetary systems. Examples of planets in a wide diversity of binary systems are known for solar-type stars, demonstrating that multiplicity and planet formation are not mutually exclusive \citep[e.g.,][]{raghavan06, bonavita07, kostov14}. However, not all multiple systems are equal in this regard. Visual binaries with separations tighter than 50--100\,au are much less likely to host long-lived protoplanetary disks \citep[e.g.,][]{cieza09} and the mature planets that orbit them have a markedly different mass distribution, suggesting that planet formation proceeds through a different mechanism than for wider systems and single stars \citep{duchene10}. Since \hab\ stars are selected based on the presence of a circumstellar disk, one may thus expect that they are less likely to possess close companions than a random sample of intermediate-mass stars. 

The organization of this paper is as follows: Section\,\ref{sec:stats} presents an up-to-date overview of the frequency of multiple systems among \hab\ stars, Section\,\ref{sec:other_props} discusses the distribution of key orbital parameters, and Section\,\ref{sec:disks} addresses the connection between multiple systems and disk properties among \hab\ stars. Finally, I discuss in Section\,\ref{sec:perspectives} some of the implications from these findings and outline some directions for future studies. 

\section{The multiplicity frequency of \hab\ stars} 
\label{sec:stats}

\subsection{General remarks}
\label{subsec:stats_gen}

Fully assessing the multiplicity properties of a sample of stars is a considerable challenge, that has only been fully achieved for nearby solar-type field stars \citep{duquennoy91, raghavan10}. Given the extremely broad distribution of binary separations, thorough multiplicity surveys must combine several observing methods, each with its inherent limitations and selection biases. Furthermore, even with a comprehensive multi-technique approach, some ranges of separations can remain unexplored as a consequence of the relatively large distances to many targets. Consequently comparisons between various multiplicity surveys are often imperfect and limited to subsets of the parameter space, and/or require a simple parametrization of the underlying distribution of physical parameters \citep[e.g.,][]{kouwenhoven07, kraus11}.

\begin{figure}[ht]
\includegraphics[width=\columnwidth]{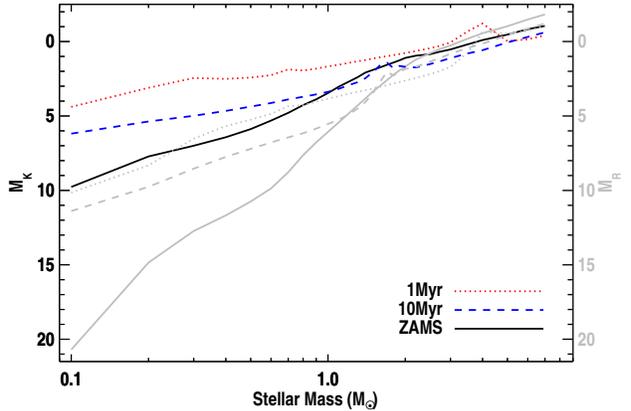}
\caption{Mass-luminosity relationships for PMS and ZAMS stars based on the evolutionary models of \cite{siess00}. The $R$ and $K$ absolute magnitudes are shown with gray and colored lines, respectively.\label{fig:mass_lum}}
\end{figure}

These difficulties are exacerbated in the case of \hab\ stars. First of all, the scarcity of intermediate-mass stars imposed by the stellar initial mass function (IMF) requires lumping \hab\ stars spanning a broad range of stellar masses into a single sample\footnote{Throughout this manuscript, I consider objects as late as early-G stars as \hab\ stars, following the most common terminology in the literature (and in line with their stellar mass $M_\star \gtrsim 2\,M_\odot$), even though these are sometimes referred to as TTS.}, even though this could smear out important but subtle trends. Furthermore, since most nearby star-forming regions only host a handful of intermediate-mass stars, most large samples of \hab\ stars are haphazard, bias-prone, mixed bags of stars with different evolutionary stages and spanning broad ranges of distances \citep{the94, vieira03, hernandez05}. Indeed, the majority of known \hab\ stars are located at least 300\,pc away from the Sun. 

One issue that is often a severe challenge when searching for companions to intermediate-mass stars is the fact that the most unequal systems (with mass ratios $q = M_2 / M_1 \lesssim 0.1$) are characterized by large contrast ratios. As shown in Fig.\,\ref{fig:mass_lum}, this is true if one considers the mass-luminosity relationship on the zero-age Main Sequence (ZAMS), which is relevant for field intermediate-mass stars whose age is typically a few 100\,Myr. However, the situation is not nearly as severe for \hab, whose young ages ensure that their low-mass companions are still in the PMS phase. As a result, achieving a contrast of 5-6\,mag in the near-infrared is sufficient to detect the photosphere of any stellar companion to an \hab\ star. While this is a favorable circumstance, it also means that 1) determining the mass of a companion hinges on the ability to assess the age of the system, and 2) companions are likely to host their own circumstellar disk which can contribute significantly to its near-infrared brightness. At visible wavelengths, where thermal emission from the disk is negligible, the contrast between \hab\ stars and their low-mass companions remains prohibitively high, even at young ages. In summary, while detecting companions to \hab\ stars is easier than for their older field counterparts, accurately characterizing them remains challenging. 

\subsection{Existing multiplicity surveys}

\subsubsection{Visual binaries}

Over the last two decades, surveys for visual binaries among \hab\ stars have achieved continuously increasing resolution and contrast, as imaging techniques improved from direct imaging \citep{pirzkal97, doering09} and speckle interferometry \citep{leinert97} to adaptive optics \citep{bouvier01, thomas07}. While sub-arcsecond companions are very likely to be bound, wider candidate companions require multi-epoch proper motion confirmation \citep[e.g,][]{hornbeck12}. The most extensive and deepest survey to date has been conducted with adaptive optics by Thomas et al. (in prep.). Expanding the survey of \cite{bouvier01}, their analysis includes 142 targets, a sample 3--5 times larger than those of \cite{pirzkal97} and \citep{leinert97}. In addition, where previous imaging searches were sensitive to companions up to 4--5\,mag fainter than their primary, the new adaptive optic surveys can detect companions up to 9\,mag fainter (albeit with a significant dependence on separation within the central 1\arcsec). 

Based on the Thomas et al. survey, the observed companion frequency for \hab\ stars is about 25\% per decade of separation in the $\approx$50--5000\,au range of projected separations. Note that this quantification is an effective way to deal with the diversity in distances to sources, since the distribution of separation of visual binaries is generally broad enough to be well approximated by a log-uniform distribution, i.e., \"Opik's law \citep{opik24}. Not surprisingly, previous surveys led to lower companion frequencies (18--20\% per decade of separation), although the modest increase suggests that the frequency of (faint) very low-mass companions is modest.

Assessing the completeness of surveys, which depends not only on the separation-dependent sensitivity of surveys but also on the underlying (unknown) distribution of mass ratios and its possible dependence on binary separation, is a challenging endeavor that is beyond the scope of the present analysis. Wider separations cannot be probed because the proper motion of most \hab\ stars is too small to safely discriminate between background stars and bona fide physical companions. Furthermore, \hab\ stars are frequently surrounded by physically associated, but unbound, low-mass PMS stars at separations of thousands of au which formed from the same parent cloud \citep{testi99}.

\subsubsection{Spectro-astrometric binaries}

Another technique to identify close visual companions is based on the detection of a spectro-astrometric signal. In a binary system comprising two stars of different effective temperature, the flux ratio of the binary varies significantly as a function of wavelength. The resulting displacements of the system's photocenter can be detected using long-slit spectroscopy. An inherent limitation of this technique is its inability to determine precisely the binary separation (only a lower limit can be robustly established), except in cases for which a complete "deblending" analysis can be performed. Broadly speaking, the spectro-astrometry method is sensitive to companions as close as $\approx$0\farcs1 and out to 2--5\arcsec\ (similar to the range probed with adaptive optics imaging, for instance), depending on seeing conditions and binary flux ratio. The companion frequency listed below applies within this approximative range.

Because it relies on spectral differences between the two components, this method is well adapted to the search of low-mass companions to \hab\ stars. Indeed, the spectro-astrometric method, which has targeted about 60 \hab\ stars to date, has revealed a companion frequency for \hab\ stars as high as $\approx$75\% \citep{baines06, wheelwright10}. This higher companion frequency than that found in direct imaging surveys may be the result of different sensitivity limits to low-mass companions. However, some candidate spectro-astrometric companions may be spurious as jets/outflows launched by the \hab\ star can also produce similar signatures. The suggestion that AB\,Aur is a binary system \citep[e.g.,][]{baines06} whereas no other observing technique has found an actual stellar companion to that source \citep{liu05, perrin09, hashimoto11} serves as a cautionary tale.

\subsubsection{Spectroscopic binaries}

While a number of spectroscopic binaries are known among \hab\ stars, systematic surveys remain few and far in between. Indeed, the combination of fast rotation, small number of photospheric features and strong emission lines is not particularly amenable to this technique. The incompleteness level of spectroscopic surveys depends on the complex interplay between the binary orbital period and mass ratio, the rotational velocity of the primary and the specific time sampling of the survey. \cite{corporon99} estimated that their survey likely missed over half of all existing spectroscopic binaries in their sample, more than making up for any possible binary-favoring bias. It is worth stressing, however, that this estimate is highly dependent on currently untestable assumptions.

With these caveats in mind, \cite{corporon99} found an observed companion frequency of about 30\%. Within a restricted range of orbital period where reasonable completeness can be ensured ($P \leq 100$\,d), they derived a minimum companion frequency of 10\%. This is in line with the binary frequency estimated by \cite{alecian13}, whose survey was not primarily designed for a multiplicity study and that had a very limited monitoring (for instance, only 11 out of their 70 targets have been observed more than twice). Thus, the companion frequency derived by \cite{corporon99} is probably a more representative estimate.

Among spectroscopic binaries, the most remarkable systems are eclipsing binaries, which offer a unique chance of accurately determining both their mass and radius, thus providing critical tests for evolutionary models. The only well-studied such system is TY\,Cra, which actually is part of a compact triple system in which the eclipsing pair consists of a 1.5 and a 2.8\,$M_\odot$ stars, respectively \citep{kardopolov81, corporon94}. Two more candidate systems have been proposed, T\,Ori and MWC\,1080 \citep{grankin92, shevchenko94a, shevchenko94b, corporon99}, although no precise determination of the stellar parameters has been published to date.

\subsubsection{Intermediate separation binaries}

Spectroscopic binaries can only probe companions out to $\approx$1\,au given the long orbital period and small orbital velocity of wider systems. However, both the imaging and spectro-astrometric techniques are sensitive to companions whose separation is at least a few tens of au. This leaves a large "separation gap" in which the search for stellar companions can only be achieved with interferometric techniques: sparse aperture masking on large monolithic telescopes and long-baseline interferometry.

While no dedicated multiplicity survey has been conducted with either technique, many \hab\ stars have been ideal targets for both of them thanks to their intrinsic brightness. A literature search indicates that 56 \hab\ stars have been observed with at least one of the two techniques, resolving one triple system \citep[GW Ori,][]{berger11} and four binary systems \citep[MWC\,361, V892\,Tau, V921\,Sco, AK Sco;][Anthonioz et al., in prep.]{millan01, smith05, kraus12}. The resulting 11\% companion frequency should be considered as a lower limit given that the contrast afforded by interferometric techniques is generally modest and that only a handful of sources have been studied with both monolithic and long-baseline interferometric methods. Furthermore, faint stellar companions can be hard to distinguish if they lie at projected separations that are commensurate with the inner regions of the circumstellar disk, as the interferometric signatures of both features are interwoven. 

\subsection{Towards a complete picture}

Although each of the survey methods discussed above has known limitations and (potentially insidious) selection biases, they nonetheless provide a nearly complete view of the multiplicity of \hab\ stars, at least out to separation of $\approx 5000$\,au. The overall companion frequency of \hab\ stars is at least 90\%, with the caveat that this estimate does not include any of the candidate spectro-astrometric companions that has not been confirmed by other methods. Considering the incompleteness of spectroscopic and interferometric surveys and the fact that field intermediate-mass stars host companions at separations as large as 45,000\,au \citep{derosa14}, it is most likely that there is at least one companion for each \hab\ star. 

Such a high multiplicity frequency may explain why many \hab\ stars are strong X-ray emitters despite the fact their internal structure should not support the existence of a coherent stellar magnetic field \cite[e.g.,][]{zinnecker94}. Indeed, it has long been proposed that the presence of a magnetically active, lower mass companion could account for this unexpected X-ray emission. However, the jury is still out as to whether this scenario applies to all cases or only a subset of the X-ray-detected \hab\ stars \citep{stelzer09}. In the latter case, the process leading to X-ray emission in single \hab\ stars remains to be identified.

\subsection{Comparison with other populations}

As mentioned above, it is highly valuable to compare the results of multiplicity studies of \hab\ stars discussed above to those of relevant stellar populations. Given the nature and scope of existing surveys, this is an exercise that is best performed by parts, however.

\subsubsection{Spectroscopic binaries}

There are too few spectroscopic binaries among \hab\ stars for a detailed analysis; only their overall frequency is reasonably well known at this point. The observed frequency of companions on short orbits (separation $\lesssim 1$\,au) is comparable to that observed for intermediate-mass stars in the field \citep{abt83}. It also is higher than the corresponding frequency among TTS \citep{melo03, nguyen12}, a dependency on stellar mass that is also observed among field stars. Thus current observations do not show significant deviations from expectations based on other stellar populations.

\subsubsection{Visual binaries}

The first surveys for visual binaries among populations of TTS revealed that wide binaries are more common in some star-forming regions than in others and, crucially, than is observed among field stars of similar masses \citep[][and references therein]{duchene99}. The population of \hab\ stars, which was poorly characterized for a long time, can now provide new insights on this topic. 

\begin{figure}[ht]
\includegraphics[width=\columnwidth]{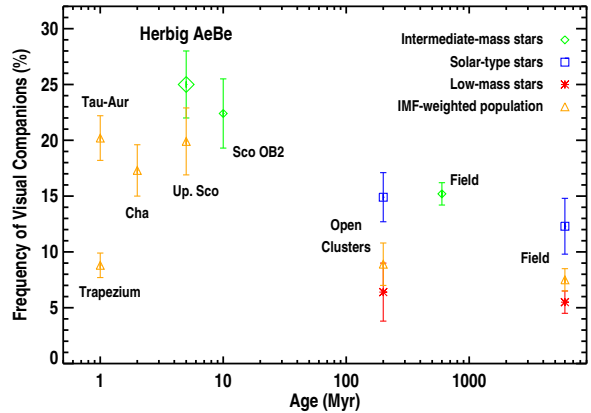}
\caption{Frequency of visual companion per decade of separation as a function of stellar age and mass. The frequencies are estimated for companions outside of $\sim$10\,au (50\,au in the case of \hab\ stars and the Orion Trapezium), over a 1- to 2-decade-wide separation range depending on the sensitivity of existing surveys. The IMF-weighted populations are based on generic surveys of PMS stars for young populations ($\lesssim 10$\,Myr) and on an indicative 70/30 split between low-mass and solar-type stars for older populations (0.1--0.5 and 0.6--1.5\,$M_\odot$, respectively). The companion frequencies shown here come from \cite{bouvier97, bouvier01b, delfosse04, kouwenhoven05, kraus08, kraus11, lafreniere08, reipurth07, reid97, derosa14, raghavan10}, Thomas et al. (in prep.) and the latest results from the RECONS survey (http://www.recons.org). The ages used for open cluster and field populations are only representative as targets typically span a relatively wide age range within each category.
\label{fig:csf_vb}}
\end{figure}

Fig.\,\ref{fig:csf_vb} illustrates the frequency of visual companions for separations ranging from a few tens to $\sim$2000\,au among a variety of stellar populations as a function of their age and stellar mass. The companion frequency observed for \hab\ stars is similar to that of the (non-disk-bearing) intermediate-mass stars population in the Sco-Cen\,OB association, but higher than that of intermediate-mass field stars. Taking only into account the statistical (binomial) uncertainties, this excess is significant at the 3$\sigma$ level. However, it is currently impossible to evaluate the extent to which selection biases affect this estimate. The companion frequency of \hab\ stars is also marginally higher than that of lower mass stars in young loose associations like Taurus-Auriga, Chamaeleon and Upper\,Sco, which have the highest companion frequency of all populations of TTS.

\section{Other properties of \hab\ multiple systems}
\label{sec:other_props}

Besides the overall frequency of companions, the distribution of orbital parameters (orbital period, mass ratio, eccentricity) and the relative frequency of binaries, triples and higher-order systems, are also rich diagnostics of the physical processes inherent to the formation and evolution of stellar systems. The incompleteness of current multiplicity surveys precludes statistical analyses of high-order systems among \hab\ stars, so I focus here on the distribution of orbital parameters.

\subsection{Orbital period distribution}

\begin{figure}[ht]
\includegraphics[width=\columnwidth]{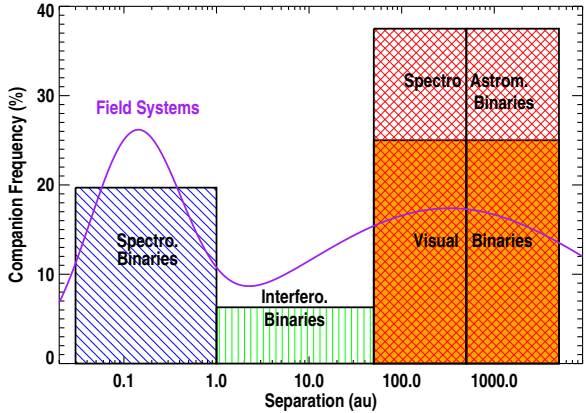}
\caption{Distribution of separation for \hab\ multiple systems. The vertical axis represents the companion frequency per decade of separation (as in Fig.\,\ref{fig:csf_vb}) over the approximate separation ranges probed by each technique, without any attempt to correct for incompleteness or overlap between them. Each technique is represented with a separate color and fill style. The purple curve is the schematic distribution for field A-type stars determined by \cite{duchene13}. \label{fig:sep_dist}}
\end{figure}

The distribution of separations for \hab\ binaries can be constructed by combining the surveys discussed above. The resulting distribution, shown in Fig\,\ref{fig:sep_dist}, confirms that \hab\ stars have a companion frequency that is similar to that of field intermediate-mass stars for short-period (spectroscopic) binaries and a significant excess for wide, visual binaries. In addition, it reveals an apparent minimum in the range probed by interferometric methods, which may also be associated with a deficit (by a factor of up to $\approx$2) relative to field stars. This comparison must be undertaken with care, as this separation range is the least well determined for field stars \citep[see, e.g.,][]{derosa14}. Furthermore, the sensitivity of interferometric surveys to stellar companions is the hardest to evaluate and plausibly the smallest among all techniques. Nonetheless, the reality of this deficit for \hab\ stars is further reinforced by the high companion frequency over this separation range for diskless B-type stars in the Sco-Cen association \citep{rizzuto13}. In summary, unless more than half of all stellar companions to \hab\ stars in that range remain undetected with interferometric techniques, both the minimum in the \hab\ separation distribution and the deficit of intermediate-separation systems relative to field stars are likely to be real. 

\subsection{Mass ratio distribution}

As discussed in Section\,\ref{subsec:stats_gen}, estimating masses for companions to \hab\ stars is not a straightforward matter. As a consequence, the mass ratio of each system is subject to significant uncertainties and comparing distributions derived from different methods may be fraught with systematic biases. It is nonetheless interesting to note that both the photometry-based distribution for visual binaries \citep{bouvier01} and the spectroscopy-informed distribution for spectro-astrometric binaries \citep{wheelwright10} are reasonably consistent with one another. In short, the observed distributions are strongly inconsistent with an IMF-pairing of components in binary systems, but instead consistent with a roughly flat mass ratio distribution, at least down to $q\approx0.2$--0.4, below which incompleteness is important. This is in line with observations of many populations of field stars \citep{duchene13}, as well as of the populations of intermediate-mass spectroscopic and visual binaries in the Sco-Cen OB association \citep{kouwenhoven05, kouwenhoven07} and those of TTS multiple systems \citep[e.g.,][]{kraus11}. Thus \hab\ binaries do not stand out among other populations of multiple systems as far as their mass ratio distribution is concerned.

\subsection{Eccentricity distribution}

\begin{figure}[ht]
\includegraphics[width=\columnwidth]{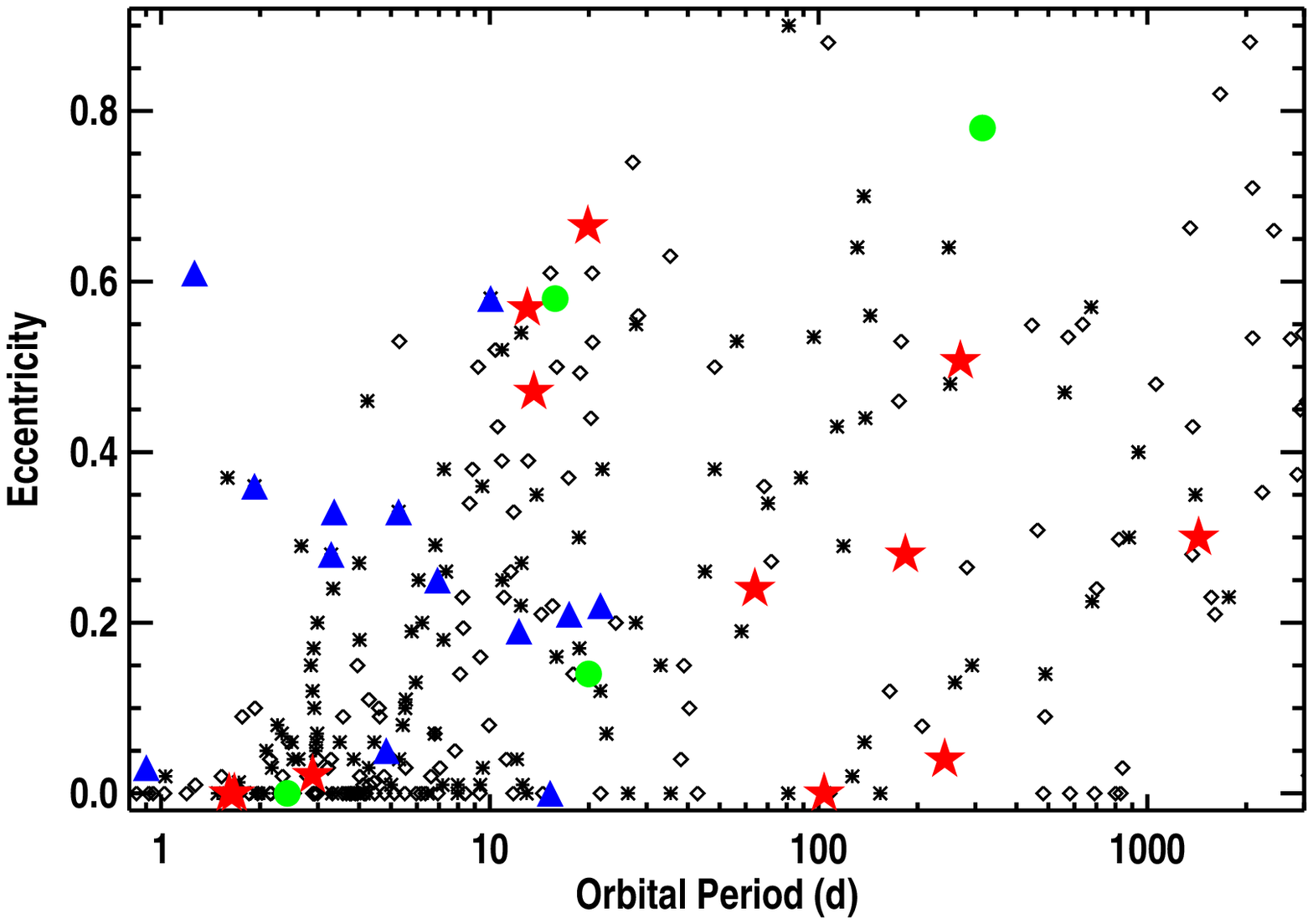}
\caption{Period-eccentricity diagram for spectroscopic binaries among \hab\ stars \citep[filled red stars;][]{mathieu91, corporon96, corporon99, alencar03, pogodin06, bohm04, bohm09, alecian09,  benisty13, beskrovnaya13}, disk-bearing TTS \citep[filled green circles; from the catalog of][]{ismailov14} and non-disk-bearing B3- through B9-type stars in the Sco-Cen OB association \citep[filled blue triangles;][]{levato87, brown97}. The black diamonds and asterisks represent populations of field A- and B-type stars, respectively \citep{abt05}.\label{fig:ecc_dist}}
\end{figure}

The number of spectroscopic binaries among \hab\ stars is limited, and only a subset of these have had their orbit estimated. While this precludes any thorough statistical analysis of the overall distribution of eccentricities for this population, it is still informative to place all \hab\ binaries with published orbits in a period-eccentricity diagram, which has been extensively studied in the past \citep[e.g.,][]{abt05}. To this end, I have compiled a list of 12 published spectroscopic orbits for \hab\ binaries that is complete to the best of my knowledge. Fig.\,\ref{fig:ecc_dist} presents the resulting period-eccentricity diagram, along with those of disk-bearing TTS and non-disk-bearing B-type stars in the Sco-Cen OB association.

Broadly speaking, \hab\ binaries span similar distributions as field A- and B-type stars. Specifically, for orbital periods longer than 10\,d, eccentricities spanning most of the (0..1) range are found, similar to disk-bearing TTS. Only three \hab\ binaries have shorter periods and all have near-circular orbits; the slightly non-zero eccentricity of the tight TY CrA\,AB pair is most likely a consequence of three-body interactions in this compact triple system \citep{beust97}. Give the scarcity of short period binaries, we can only conclude that the circularization period for \hab\ systems is in the 3--10\,d range, similar to that of field A-type stars \citep[e.g.,][]{abt05}. In turn, this suggests that circularization in intermediate-mass systems occurs on a timescale shorter than the typical age of \hab\ stars, hence $\lesssim$10\,Myr. This may be faster than previously believed \citep{abt02}, although uncertainties associated with tidal dissipation mechanisms remain large \citep[e.g.,][]{beust97}. Interestingly, 1--10\,d B-type binaries in the Sco-Cen OB association have significantly higher eccentricities than \hab\ systems (see Fig.\,\ref{fig:ecc_dist}) hinting at a significant mass-dependency for the tidal circularization, or at a causal link between disk survival and eccentricity of the central binary (i.e., a long-lived disk is able to circularize the binary orbit).

\section{\hab\ stars, multiplicity and disks}
\label{sec:disks}

With so many known \hab\ binaries, it is now possible to investigate the influence of multiplicity on their associated disks on empirical grounds. Here I discuss several key properties of \hab\ disks in the context of binary systems and, whenever possible, compare any trend with the situation of disks in lower-mass TTS binary systems.

\subsection{Disk-companion interactions}

\subsubsection{Influence of the binary separation}

First of all, the high frequency of companions observed among \hab\ stars is a clear confirmation that multiplicity and circumstellar disks are not mutually exclusive. However, the observed distribution of separations (Fig.\,\ref{fig:sep_dist}) suggests that companions at intermediate separations, roughly in the 1--50\,au range, are less frequent among \hab\ stars than the overall population of intermediate-mass stars, a conclusion reminiscent of the results for TTS \citep{cieza09, kraus12b}. 

Given the breadth of the binary separation distribution and the typical size of protoplanetary disks ($\sim 100$\,au), \hab\ disks can be split into two distinct categories: circumstellar when the companion is a distant one, or circumbinary in the case of a tight pair. This diversity is also observed for TTS \citep[e.g.,][]{harris12} as well as debris disks around solar-type and intermediate-mass MS stars \citep{trilling07, rodriguez12}. As may have been expected, the presence of a companion to an \hab\ object therefore seems to have a very similar influence on its disk than for lower-mass primaries.

\subsubsection{\hab\ and transition disks}

Most circumbinary disks surround close, spectroscopic binaries. As pointed out above, slightly wider companions tend to completely disrupt the disk rather than simply perturbing it. There are exceptions to this rule, however, as revealed by the examples of the TTS systems GG\,Tau and UY\,Aur, for instance \citep[e.g.,][]{roddier96, close98}. Such systems may play an important role in the so-called "transition disks" phenomenon. Those were first identified among TTS as disks presenting massive mid- and far-infrared excesses but essentially no near-infrared excess \citep{najita07, espaillat14}. This indicates that the innermost (hottest) regions of the disk have been cleared of dust, leaving only warm and cold dust further out. Several mechanisms can be responsible for this situation but one of them is the presence of a close (sub)stellar companion at a separation of a few au, as demonstrated in the cases of CoKu\,Tau\,4 and LkCa\,15 \citep{ireland08, kraus12c}. 

Since transition disks appear to be a common occurrence among \hab\ stars as well \citep[e.g.,][]{maaskant13, yasui14}, it is natural to wonder whether some of the large holes observed around young intermediate-mass stars are carved by low-mass companions. Despite numerous searches for companions in \hab\ transition disks, the only confirmed such case to date is HD\,142527, where a low-mass stellar companion has carved a large gap between the inner and outer regions of the disk \citep{biller12, close14}. It is therefore likely that the formation of a gap/hole in the inner regions of \hab\ disk is only rarely driven by the presence of a stellar companion, a similar conclusion as for TTS systems. 

The focus has shifted in recent years toward planetary-mass companions, which are now accessible thanks to improved contrast capabilities. A planetary mass companion has been proposed in the disk surrounding HD\,100546 although it does not lie within a dust-empty region \citep{quanz13, avenhaus14}. More recently, a similar companion has been proposed to reside within the inner gap of the HD\,169142 disk \citep{biller14, reggiani14}. Further observations are required to confirm the nature of this object and, more broadly, to test the hypothesis that newly-formed planets are responsible for the inner hole of transition disks. 

\subsection{Disk-orbit relative orientation}

Depending on the formation scenario of the binary system, the disk midplane and the orbital plane can be tilted relative to one another. Therefore, the relative orientation of the disk and orbital planes is an important clue about the formation process of the system. Unfortunately, there are very few \hab\ binary systems in which both the orbit and disk orientation can be ascertained, the main limitation usually being the precise characterization of the orbital motion. Still, some \hab\ systems are amenable to individual studies thanks to their unique properties, as discussed below. Here I consider separately the cases of wide and close binaries, as they are probed via different methods.

\subsubsection{Circumstellar disks in wide binaries}

Most \hab\ visual binaries have orbits that are far too long to be derived, so it is usually not possible to determine the relative orientation of the orbit and the disk. There are however remarkable exceptions to this general rule. For instance, both the outflow and the binary orbit of the LkH$\alpha$\,198 system lie close to the plane of the sky, indicating that the circumstellar disk is nearly orthogonal to the binary orbit \citep{smith05}. A strong argument can also be made in the PDS\,144 system, where one of the two components has its disk viewed almost exactly edge-on while the other lies at a lower inclination, indicating that the two disks are misaligned by about 25\degr\ \citep{perrin06, hornbeck12}. Necessarily, at least one of the two disks is misaligned with the (currently unknown) orbital plane. 

\begin{figure}[ht]
\includegraphics[width=\columnwidth]{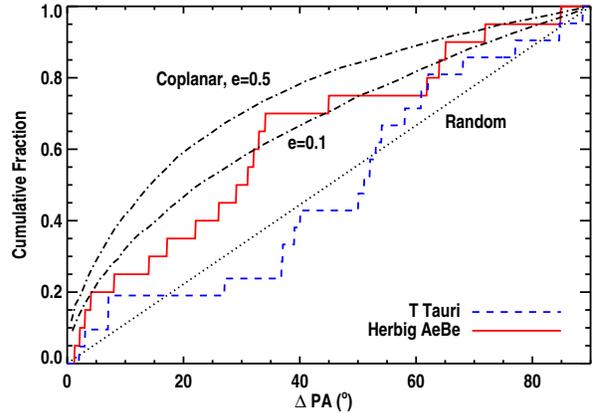}
\caption{Relative alignment of circumstellar disks in \hab\ and TTS wide binary systems (red and blue curves, respectively). The measured angle ($\Delta$PA) is the difference in position angle between the disk semi-major axis and the projected separation of the binary system. The distribution for \hab\ systems is from \cite{wheelwright11} while data for TTS systems are from high-resolution millimeter imaging surveys \citep{andrews07, andrews10, guilloteau11, harris12, akeson14} after exclusion of systems in which the disk is too close to pole-on ($i \geq 30$\degr) or whose position angle is not known to better than 20\degr. The dotted and dot-dashed curves represent the expected distributions if the disk and orbital planes are randomly oriented and perfectly coplanar \citep[for fixed eccentricities of $e=0.5$ and $e=0.1$;][]{wheelwright11}, respectively. \label{fig:disk_align}}
\end{figure}

Beyond individual systems, the only large-scale analysis of disk orientation to date has relied on spectro-polarimetric determination of disk orientation and on a statistical treatment of the projection effects of the binary orbit and of its eccentricity. Using this method, \cite{wheelwright11} concluded that disks are not randomly oriented relative to their orbit (albeit only at the 2.2$\sigma$ confidence level), but instead that they are generally aligned with the binary orbital plane. However, it must be noted that this conclusion rests on the assumptions that most wide systems have low eccentricities. Instead, wide binaries among intermediate-mass field stars have an essentially uniform distribution of eccentricities, with an average eccentricity $\overline{e} \approx 0.5$ \citep{abt05}. If this also applies for \hab\ systems, the observed distribution of disk/orbit misalignment angles suggests that their disks only have a modest degree of alignment with the binary orbital plane (Fig.\,\ref{fig:disk_align}), i.e., the average angle between the orbital and disk plane in \hab\ binaries is small but non-zero (say, $\overline{\Delta i} \lesssim 30$\degr).

While this is reminiscent of the conclusion that disks in TTS binary systems tend to be aligned with one another \citep[][and references therein]{monin07}, an up-to-date distribution of disk-orbit misalignment angles based on recent high-resolution millimeter imaging surveys is most consistent with a completely random orientation for these disks (see Fig.\,\ref{fig:disk_align}). Furthermore, as for \hab\ systems, strongly misaligned disks have been observed in wide TTS binaries \citep[e.g.,][]{jensen14}. Overall, protoplanetary disks in low-mass binaries appear even more randomly oriented relative to their system's orbit than among \hab\ systems.

\subsubsection{Circumbinary disks around tight binaries}

While their orbits are conveniently short, spectroscopic orbital solutions are usually marred by an inclination ambiguity. This problem can be solved in cases where the system can be spatially resolved (typically through long-baseline interferometric observations). Detailed analyses have been possible in case of the GW\,Ori, HD\,200775 and AK\,Sco systems \citep[][Anthonioz et al., in prep.]{vink05, berger11, benisty13}. In all three cases, it has been found that the disk, whose inner edge is located outside of the binary system, is (nearly) coplanar with the orbit. A possible counter-example is the V380\,Ori system, where the spectroscopic binary has its orbital plane viewed pole-on \citep{alecian09}, while the giant outflow from the system propagates more or less in the plane of the sky \cite{reipurth13}, hence the corresponding disk would be close to edge-on. However, it is possible that the outflow in this system arose from a violent dynamical reconfiguration, in whih case any directionality may not be indicative of the long-term disk orientation. Furthermore, it is not obvious that an outflow originating from a circumbinary disk should be exactly orthogonal to the latter, as the interaction with the inner system could easily break any symmetry. In any case, it appears that coplanarity for circumbinary disks is the rule in tight (spectroscopic and interferometric) \hab\ binaries. 

Circumbinary disks among TTS systems display a similar degree of coplanarity, as illustrated by the cases of UZ\,Tau\,E and V4046\,Sgr \citep{prato02, jensen07, rosenfeld12}. Interestingly, the widest TTS binary to host a circumbinary ring, the 35\,au-separation GG\,Tau, has an orbit that is tilted by 15--25\degr\ relative to the plane of the disk \citep{beust05, kohler11}, suggesting that near-perfect coplanarity only applies to systems whose semi-major axis does not exceed a few au.

\section{Implications and perspectives}
\label{sec:perspectives}

As I have shown in Section\,\ref{sec:stats}, the initial assumption of a very high multiplicity of \hab\ stars is borne out by observations. With an average of at least one companion per \hab\ star, it is clear that multiplicity should not be considered a relatively rare phenomenon that can generally be ignored when studying an object in depth. Instead, low spatial resolution observations of objects whose multiplicity status has not yet been assessed should be interpreted with care. This is especially true of observations taken at wavelengths at which even a lower-mass companion can contribute significantly to the system's brightness, such as the far-infrared and millimeter ranges where disks around TTS companions can be quite bright. This may even be a factor to consider for future mid-infrared {\it JWST} observations of \hab\ stars. However, this high degree of multiplicity should not be interpreted as evidence that all \hab\ stars are members of binary systems either, as high-order multiple systems compensate for the small, but finite, fraction of single stars, akin to the situation of TTS \citep{kraus11}. 

The main multiplicity properties of \hab\ stars, i.e., the companion frequency and distributions of orbital parameters, are mostly consistent with those observed for similarly young, but non-disk-bearing, intermediate-mass stars in the Sco-Cen OB association, indicating that the \hab\ phenomenon is not very sensitive to the presence of a companion. While \hab\ stars host more companions overall than the lower-mass TTS, the two populations share roughly flat mass-ratio and eccentricity distributions (for $P \gtrsim 10$\,d), suggesting that the star formation process proceeds through similar mechanisms over the entire 0.1--8\,$M_\odot$. 

The apparent excess of visual companions for \hab\ stars over their MS counterparts is intriguing. Among populations of TTS, a similar excess has long been observed, and the debate as to whether this excess is indicative of intrinsic differences in the star formation process itself, or a consequence of intense dynamical evolution in young stellar clusters is still ongoing \citep[e.g.,][]{king12, marks12}. In any case, the excess has been interpreted as evidence that a majority of field stars form in dense clusters, in which wide companions are much less common \citep[e.g.,][]{patience02, kohler06, reipurth07}. It is unclear whether a similar line of thought can be applied to intermediate-mass stars, as \hab\ stars are distributed among both scattered and clustered populations, and thus do not represent a unique star-forming environment. The fact that intermediate-mass stars in the Sco-Cen association have a similar fraction of wide companions may instead suggest that this high companion frequency is universal. In turn, this would imply that a subset of these initial companions are in unstable configurations and are dispersed on a timescale of $\gtrsim$10\,Myr. Probing the frequency of wide companions in populations of intermediate-mass stars in young open clusters ($\lesssim$100\,Myr) would go a long way toward understanding this evolution. Unfortunately, the speckle interferometry survey of the $\alpha$\,Per cluster by \cite{patience02} did not have sufficient sensitivity to low-mass stellar companions to be conclusive. New surveys using the high-contrast capabilities of planet-searching instruments (e.g., GPI, SCExAO, SPHERE) will provide a decisive input to this question.

As for lower mass PMS objects, the presence of a stellar companion does not appear to have dramatic effects on the circumstellar disks surrounding \hab\ stars, with the likely notable exception of companions at intermediate separations ($\approx$1--50\,au). This can be readily understood as a companion in that separation range could dynamically disrupt any disk, or even prevent its formation altogether. It is worth pointing out, however, that when disks are present around Myr-old TTS  in such binary systems, their lifetime is essentially the same as that of wider pairs and single stars \citep{kraus12b}. Therefore, disks found in intermediate separation \hab\ binaries probably offer similar prospects to forming planetary systems as those around single stars, except that they are less common to begin with. In this context, it is worth restating that the transition disk phenomenon is in most cases not related to multiplicity. Instead, it is likely that the formation of a gap/hole in the inner regions of \hab\ disk is a consequence of similar disk evolution processes as for lower mass TTS. The combination of ALMA sub-mm observations with high contrast scattered light images is the most promising approach to understand the nature of these systems.

The diversity of architectures for disks in \hab\ multiple systems suggests that planet formation around intermediate-mass stars can lead to both circumstellar and circumbinary planets, as is observed for solar-type stars. All planetary systems known among intermediate-mass stars have a circumstellar architecture but this is a consequence of the fact that they have only been searched via the radial velocity method \citep{johnson10}, whose precision is significantly limited in the case of close binary systems \citep{konacki09}. As the {\it Kepler} mission has shown, circumbinary planets are much easier to detect via the transit method, but the latter has not yet been employed much in the context of intermediate-mass stars. Nonetheless, it is natural to expect that such systems exist and will be discovered in the future, for instance as part of the K2 phase of the {\it Kepler} mission.

Finally, the tentative near-perfect coplanarity of circumbinary disks around close, spectroscopic binaries (both among \hab\ and TTS) is in line with the configuration of the {\it Kepler}-discovered circumbinary planets around tight solar-type binaries \citep{kostov14}. Astrometric monitoring of the orbit of spectroscopic \hab\ binaries with GAIA will likely increase manifold the number of systems in which this coplanarity can be tested. If confirmed, this coplanarity suggests a disk fragmentation origin for close binaries, although it is plausible that tidal torques can force at least the innermost region of the disk to settle in the same plane as the binary orbit if it is initially misaligned. Among wider systems, instead, circumstellar disks appear to be only moderately aligned with the orbital plane. This conclusion is reminiscent of the architecture of triple stellar systems \citep{hale94}. Turbulent fragmentation of the parent cloud is the leading mechanism to generate such a configuration. In this scenario, the memory of the orientation of the cloud's angular momentum vector may only be partially erased from the various fragments. Furthermore, the fact that disks in TTS binaries are less aligned than those of \hab\ systems may indicate that the higher stellar masses of the latter are able to generate significant torques over timescales of a few Myr. Observations of the relative orientation of disks in the youngest (embedded) intermediate-mass binaries in the future would help in determining the exact initial configuration of the newly formed systems.

\acknowledgments
Sandrine Thomas, Bernadette Rodgers, Fabien Anthonioz, J\'er\^ome Bouvier, Adam Kraus, Rene Oudmaijer and an anonymous referee provided me with valuable feedback on various aspects of this work and shared some of their results in advance of publication, which I greatly appreciate. I am also grateful to the editors of the Topical Collection on \hab\ stars (Willem-Jan de Wit and Rene Oudmaijer) for inciting me to delve into this rich topic.

\nocite{*}
\bibliographystyle{spr-mp-nameyear-cnd}
\bibliography{gduchene}

\end{document}